\documentclass[acmsmall,screen]{acmart}

\usepackage{tikz}

\usepackage{enumitem}
\usepackage{xcolor}
\usepackage{tabularx}
\usepackage{booktabs}
\usepackage{array}
\usepackage{calc}
\usepackage{multirow}
\usepackage{wrapfig}
\usepackage{arydshln}  
\usepackage{subcaption}  

\definecolor{softpink}{RGB}{219,112,147}

\newcommand{\pinkhref}[2]{%
  \href{#1}{\textcolor{softpink}{#2}}%
}

\begin{document}


\newcommand{\name}[0]{RippleGUItester}

\title[]{\name{}: Change-Aware Exploratory Testing}



\author{Yanqi Su}
\email{yanqi.su@tum.de}
\affiliation{%
  \institution{Technical University of Munich}
  \country{Germany}
}

\author{Michael Pradel}
\email{michael@binaervarianz.de}
\affiliation{%
    \institution{CISPA Helmholtz Center for Information Security}
  \country{Germany}
}

\author{Chunyang Chen}
\email{chun-yang.chen@tum.de}
\affiliation{%
  \institution{Technical University of Munich}
  \country{Germany}
}

\begin{abstract}
  Software systems evolve continuously through frequent code changes, yet such changes often introduce unintended bugs despite extensive testing and code review. 
Existing testing approaches are largely constrained to predefined execution paths or rely on unguided exploration, leaving many change-induced issues undetected. 
To address this challenge, we present \textit{\name{}}, a change-driven testing system that treats a code change as the epicenter of a ripple effect and explores its broader, user-visible impacts via the GUI.
Given a code change, \name{} performs LLM-based change-impact analysis to generate and enrich realistic test scenarios,
executes these scenarios on both pre-change and post-change versions of the system, and applies differential analysis to identify behavioral differences.
Crucially, \name{} employs multimodal bug detection, comparing visual GUI changes and interpreting them in the context of natural-language change intents to distinguish unintended bugs from intended behavioral updates.
We evaluate our approach on hundreds of real-world code changes across four widely used software systems: Firefox, Zettlr, JabRef, and Godot.
Our results show that the proposed approach uncovers bugs introduced by code changes that were missed by existing test suites, CI pipelines, and code review. 
In total, we identify 26 previously unknown bugs that still exist in the latest versions of the evaluated systems.
After reporting,
16 bugs have been fixed, 2 have been confirmed, 6 are still under discussion, and 2 were marked as intended.
We envision \name{} being applied before or shortly after a code change is merged, enabling earlier detection of regressions.


\end{abstract}





\maketitle

\section{Introduction}Modern software systems evolve continuously through frequent code changes. 
While such changes are essential for adding features, fixing defects, and improving maintainability, 
they also risk introducing new bugs. 
Despite extensive testing and code review, a non-negligible fraction of code changes still introduce new bugs~\cite{bug_introduction_mostafa2017experience,bic_kamei2012large,bic_kim2008classifying,bic_sliwerski2005changes,bic_yin2011fixes,bic_wen2019exploring}, 
negatively affecting user experience and incurring substantial debugging and maintenance costs.


To better understand this phenomenon, we analyze code changes and the bugs they introduce.
Even with rigorous testing, 12.2\% of analyzed code changes introduced new bugs, highlighting the prevalence of this problem in real-world software development.
Fig~\ref{fig:motivation_example} shows a representative example from the Firefox browser.
When fixing 
\pinkhref{https://bugzilla.mozilla.org/show_bug.cgi?id=1858633}{Issue~1858633} 
(\textit{Passwords page opens with incorrect URL fragment}), which is triggered by opening the Passwords page from the App Menu while a site is open, the code change unintentionally introduced another bug (\pinkhref{https://bugzilla.mozilla.org/show_bug.cgi?id=1937085}{Issue~1937085}).
This new bug is triggered when the Passwords page is accessed from a pop-up window, causing it to open in the pop-up rather than the main browser window.
This example highlights several key challenges in detecting bugs introduced by code changes:
(i) diverse triggering event sequences, such as accessing the same page through a pop-up window instead of the App Menu;
(ii) difficulty in constructing appropriate test data, for instance requiring a website that can trigger a login pop-up window (e.g., https://x.com/); and
(iii) hard-to-predict test oracles for GUI behavior, such as whether the Passwords page opens in the main window.


Bugs may also arise from cross-scenario side effects, where fixing one usage scenario inadvertently disrupts another seemingly unrelated one.
For example, fixing a missing scrollbar on the Settings page (\pinkhref{https://bugzilla.mozilla.org/show_bug.cgi?id=1792881}{Issue~1792881}) unintentionally caused the search box
to shift when no search results were returned (\pinkhref{https://bugzilla.mozilla.org/show_bug.cgi?id=1793730}{Issue~1793730}).
Although both issues occur within the Settings page, they are triggered by different usage scenarios, scrolling versus performing a search, making such side effects difficult to anticipate during testing.
Thus, many bugs escape existing testing procedures, such as regression testing and automated CI pipelines, which are largely constrained to predefined execution paths and expected behaviors. 
Exploratory testing~\cite{exploratory_tutorial,exploratory_nature,exploratory_book_tips_tricks_tours_techniques_to_guide_test_design}, while effective at uncovering unexpected issues, is not driven by code changes and provides little systematic guidance on what to explore after a change.
This gap motivates the need for a testing approach that is explicitly driven by code changes yet capable of exploring their broader effects.

This work presents \textit{\name{}}, a change-driven testing system that aims to uncover user-visible issues induced by a code change by systematically exploring the ripple effects of that change via the GUI.
Rather than focusing solely on the intended modification, \name{} enables the discovery of indirect and cross-scenario effects that are difficult to anticipate upfront and are often missed by existing testing approaches.
\name{} consists of three components: Test Scenario Generator, Test Scenario Executor, and Bug Detector.
Starting from a given code change, the Generator performs LLM-based change-impact analysis to identify potentially affected behaviors and generate initial test scenarios.
These scenarios are enriched with diverse triggering event sequences derived from scenario knowledge mined from historical issues and pull requests.
To make scenarios executable, concrete test data is instantiated, producing realistic and fully executable test scenarios.
The Executor then runs each scenario on both the pre-change and post-change builds of the software under test.
Finally, the Detector adopts a differential analysis strategy.
It detects and annotates visual differences between GUI screenshots from the two builds and analyzes them in the context of the executed scenario and the change intent to determine whether an observed difference corresponds to an intended behavior change or an unintended bug.

Our evaluation applies \name{} to hundreds of real-world code changes from four widely used and complex software systems: \pinkhref{https://www.firefox.com/en-US/}{Firefox}, \pinkhref{https://www.zettlr.com/}{Zettlr}, \pinkhref{https://www.jabref.org/}{JabRef}, and \pinkhref{https://godotengine.org/}{Godot}.
The results show that \name{} is able to detect bugs introduced by code changes that remain unnoticed by existing test suites, CI tools, and code review.
Overall, \name{} uncovered 26 previously unknown bugs that still exist in the latest versions of the evaluated systems.
After reporting these bugs,
16 bugs have been fixed, 2 have been confirmed, 6 are still under discussion, and 2 were determined to be intended behavior.
These results indicate that \name{} is effective in uncovering unintended, user-visible behavioral changes caused by code change.
While \name{} incurs relatively high overhead (54.8 minutes and \$5.99 per pull request on average), it demonstrates the ability to uncover real-world GUI regressions that are difficult to detect automatically.

In summary, this paper makes the following contributions:
\begin{itemize}[leftmargin=*]
    \item We observe and analyze code changes and the bugs they introduce in the real-world software system, quantifying their prevalence and highlighting the practical challenges that make such bugs difficult to detect.

    \item We introduce \name{}, the first change-aware GUI testing system that uncovers user-visible issues induced by code changes by systematically exploring their ripple effects across usage scenarios.


    \item We evaluate \name{} on hundreds of real-world code changes from four widely used software systems, demonstrating its effectiveness by uncovering 26 previously unknown bugs  in the latest versions of the evaluated systems: 16 have been fixed, two have been confirmed by developers, six remain under discussion, and two were marked as working as intended.

    \item We make our dataset and implementation publicly available to foster future research (see Section~\ref{sec:data} for details). 
\end{itemize}

\section{Motivation}\label{sec:motivation}


Code changes, whether for implementing tasks, requesting features, fixing bugs or refactoring, can inadvertently introduce new bugs.
To better understand this phenomenon, we examine how often code changes result in new bugs and identify challenges that make such bugs difficult to anticipate and detect in practice.
Throughout this paper, 
we use the term \textit{Issue} to denote any task, feature, bug, or refactoring, and the term \textit{Pull Request (PR)} to denote the code changes made to fix an issue.

\subsection{Prevalence of Bug-Inducing Code Changes}
To investigate the prevalence of this phenomenon, we analyzed 97,347 PRs merged to the \pinkhref{https://www.mozilla.org/en-US/firefox/new/}{Firefox} between May 1, 2019, and January 1, 2025. 
We selected 
Firefox as the study subject due to its
comprehensive bug management system, \pinkhref{https://bugzilla.mozilla.org/home}{Bugzilla}.
This system provides traceability between PRs and both the issues they resolve and the unintended bugs they introduce.
Despite rigorous quality assurance procedures such as linting, unit testing, regression testing, and code review prior to integration into the source code repository, 11,910 PRs still introduced new bugs, accounting for 12.2\% of all changes. 
This percentage represents a conservative estimate, as some bugs are never detected and others may be detected but not linked to the bug-introducing PR.
This indicates that the phenomenon is widespread in real-world practice.
The bugs are typically discovered by users, testers, or developers in the subsequent versions. 
Evidently, this not only impacts user experience but also incurs significant testing, debugging and fixing overhead.

\vspace{1mm}
\noindent\fcolorbox{softpink}{white}{\begin{minipage}{13.7cm}
Despite rigorous testing, 11,910 PRs (12.2\% of all changes) still introduced new bugs, underscoring the widespread and significant nature of this phenomenon.
\end{minipage}}\\
\vspace{1mm}



\subsection{Challenges of Change-Induced Bug Detection}\label{subsec:characteristics}
Our analysis shows that these bugs are not simple regressions and therefore cannot be reliably detected by rerunning existing regression tests.
Instead, they often involve complex triggering steps and hard-to-predict test oracles.
We illustrate these challenges with concrete examples.

\subsubsection{Challenging Trigger Steps} 
Trigger steps consist of two key parts: event sequences and test data.
\textit{Event sequences} are the sequences of user actions or system events that drive the software into specific states.
\textit{Test data} are sets of inputs or information used to verify the correctness, performance, and reliability of software systems.
We find that the main challenge in triggering introduced bugs is the diversity of event sequences and test data involved in complex software systems.

\begin{figure}[!ht]
  \centering
  \begin{subfigure}{0.470\textwidth}
    \centering
    \includegraphics[width=\linewidth]{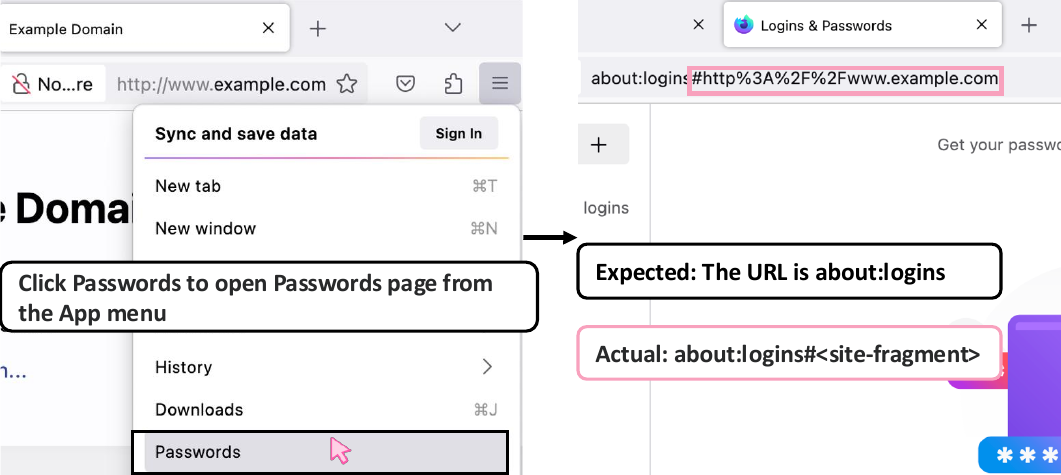}
    \vspace{-2mm}  
    \caption{\pinkhref{https://bugzilla.mozilla.org/show_bug.cgi?id=1858633}{Issue1858633}: Open with site fragment in URL}
    \label{fig:motivation_example1a}
  \end{subfigure}
  \hfill
  \begin{subfigure}{0.515\textwidth}
    \centering
    \includegraphics[width=\linewidth]{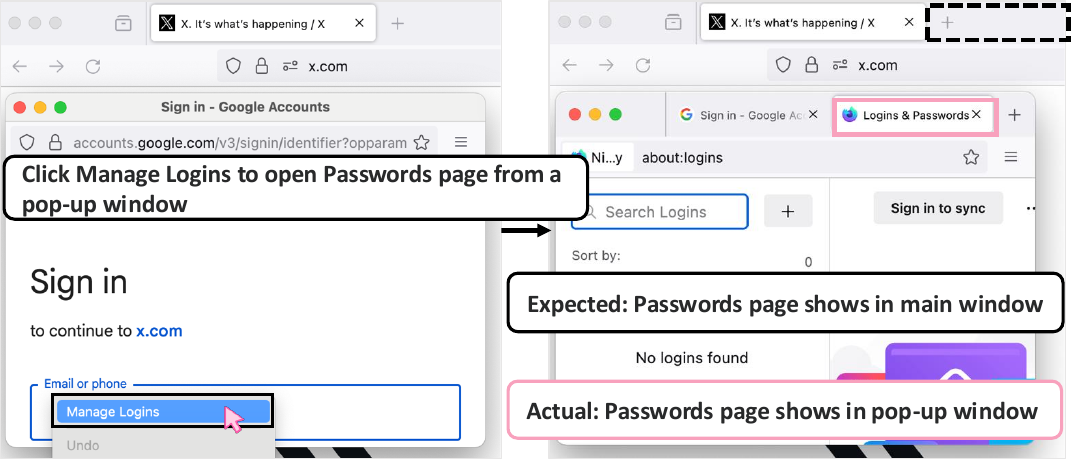}
    \caption{\pinkhref{https://bugzilla.mozilla.org/show_bug.cgi?id=1937085}{Issue1937085}: Open in pop-up window}
    \label{fig:motivation_example1b}
  \end{subfigure}
  \vspace{-2mm}
  \caption{Fixing \pinkhref{https://bugzilla.mozilla.org/show_bug.cgi?id=1858633}{Issue1858633} introduces \pinkhref{https://bugzilla.mozilla.org/show_bug.cgi?id=1937085}{Issue1937085}. 
  Only the final event is shown. }
  \label{fig:motivation_example}
  \vspace{-3mm}
\end{figure}

\textbf{Diverse Event Sequences.}
As shown in Fig~\ref{fig:motivation_example1a}, \pinkhref{https://bugzilla.mozilla.org/show_bug.cgi?id=1858633}{Issue~1858633}
(\textit{Passwords page opens with incorrect URL fragment}) is triggered when the Passwords page is opened from the App Menu while a website is open.
The PR that fixed this issue introduced another bug
(\pinkhref{https://bugzilla.mozilla.org/show_bug.cgi?id=1937085}{Issue~1937085}: \textit{about:logins opens in Sign-In popup window instead of main Firefox window}).
As shown in Fig~\ref{fig:motivation_example1b}, this bug is triggered when the Passwords page is opened from a pop-up window.
Specifically, the event sequence involves opening a website with a pop-up login window (e.g., https://x.com/), clicking the ``Sign in'' button to activate the pop-up, right-clicking the input field, and selecting ``Manage Logins''.
Thus, detecting bugs during the code change stage requires considering diverse event sequences.

\textbf{Diverse Test Data.}
\pinkhref{https://bugzilla.mozilla.org/show_bug.cgi?id=1750072}{Issue~1750072}
 introduces a built-in password reveal/conceal control for password inputs.
Testing this enhancement requires exercising the same event sequences, i.e., opening a UI containing password fields and observing the reveal/conceal control.
These different UIs containing password fields constitute distinct test data.
However, due to the large number of such UIs, exhaustive testing is impractical. 
As a result, testing overlooked the Thunderbird mail account setup page (\pinkhref{https://bugzilla.mozilla.org/show_bug.cgi?id=1913889}{Issue~1913889}: \textit{Two buttons to toggle password visibility}), and the Save Password doorhanger (\pinkhref{https://bugzilla.mozilla.org/show_bug.cgi?id=1936548}{Issue~1936548}: \textit{Redundant Show password option}).
Thus, generating representative and diverse test data is essential for adequately validating code changes.

\textbf{Cross-Scenario Side Effects.}
PRs that resolve a specific issue may also affect other usage scenarios that appear unrelated at first glance.
For example, fixing \pinkhref{https://bugzilla.mozilla.org/show_bug.cgi?id=1792881}{Issue~1792881}
 (a missing vertical scrollbar on the Settings page) unintentionally caused \pinkhref{https://bugzilla.mozilla.org/show_bug.cgi?id=1793730}{Issue~1793730}, where the search box and page content shift when no search results are returned.
Although both issues occur within the same Settings interface, they correspond to different user interaction scenarios: scrolling the settings page versus performing a search that yields no results.
When testing the fix for scrollbar issue, it is difficult to anticipate that the search-related scenario should also be tested.
This example highlights the challenge of uncovering hidden dependencies across seemingly unrelated usage scenarios.


\subsubsection{Hard-to-Predict Test Oracles}
To catch introduced bugs, it is not enough to only trigger them.
Corresponding test oracles are also required to detect their manifestations. 
However, as the examples above illustrate, the test oracles are highly irregular and hard-to-predict. 
For instance, fixing the incorrect Passwords URL led to the Passwords page opening in a pop-up window; enabling the built-in password reveal/conceal control resulted in redundant “Show password” options, and resolving the missing vertical scrollbar caused the search bar to shift unexpectedly. 
These cases demonstrate that rule-based test oracles are infeasible. 
Being able to capture such irregular and hard-to-predict oracles remains another major challenge.

\vspace{1mm}
\noindent\fcolorbox{softpink}{white}{\begin{minipage}{13.7cm}
Detecting bugs is inherently challenging because introduced bugs often require diverse event sequences, varied test data, and interactions across multiple usage scenarios to be triggered.
Moreover, their manifestations depend on hard-to-predict test oracles.
These factors together explain why many bugs escape existing testing procedures.
\end{minipage}}\\
\vspace{1mm}

\section{Approach}\label{sec:approach}


\begin{figure}

    \centering
    \includegraphics[width=1\textwidth]{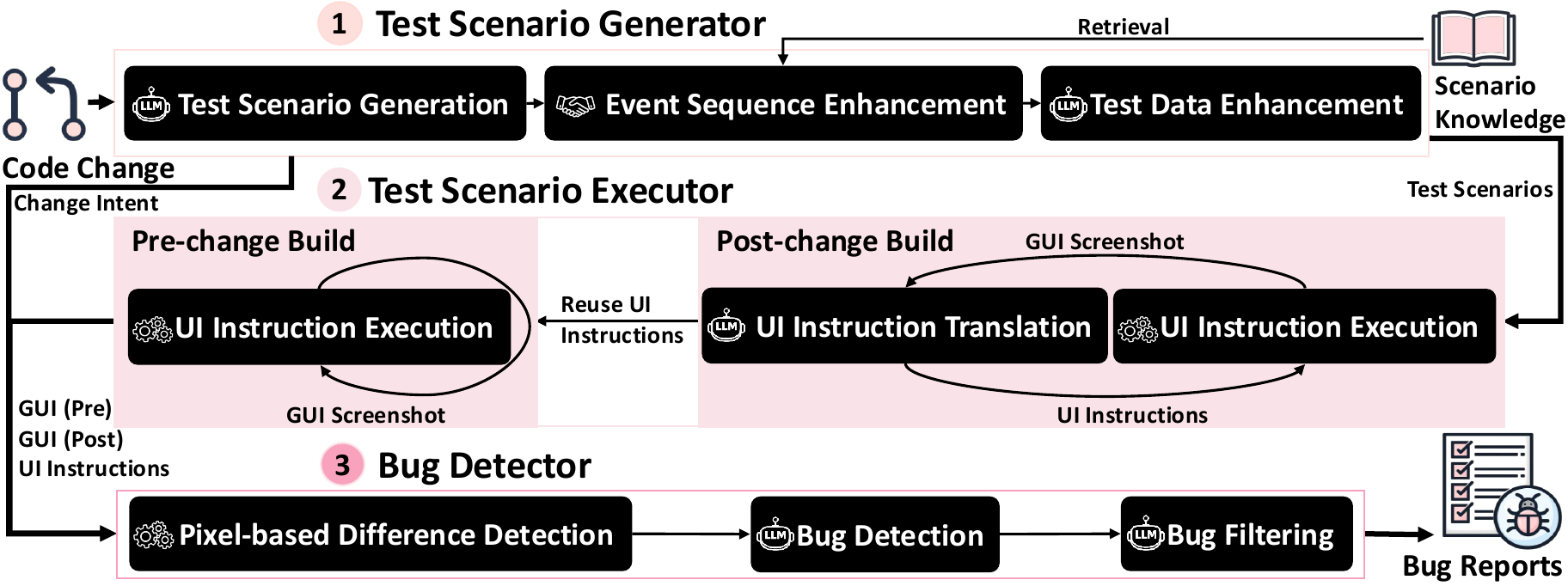}
    \caption{Approach Overview. 
    Icons denote types: LLM-based (robot), algorithmic (gear), and hybrid (handshake). 
    }
    \label{fig:approach_overview}
        \vspace{-0.45cm}

\end{figure}

To capture bugs introduced by code changes, we present \textit{\name{}}, a novel testing technique that integrates the natural language understanding, code comprehension, and vision capabilities of multimodal LLMs with application-specific knowledge.
As shown in Fig~\ref{fig:approach_overview}, the system comprises three components: Test Scenario Generator, Test Scenario Executor, and Bug Detector (hereafter abbreviated as \textit{Generator}, \textit{Executor}, and \textit{Detector}). 
The \textit{Generator} analyzes a given code change to identify potentially affected usage scenarios and generate executable test scenarios (Section~\ref{sec:generator}).
The \textit{Executor} runs these scenarios on both the pre-change and post-change builds of the SUT (Section~\ref{sec:executor}).
The \textit{Detector} compares execution results across the two builds to determine whether observed differences correspond to intended behavioral changes or unintended bugs (Section~\ref{sec:detector}).

\subsection{Test Scenario Generator}\label{sec:generator}
The Generator analyzes the potential impacts of code changes and generates corresponding test scenarios by incorporating application-specific knowledge.
As discussed in Section~\ref{subsec:characteristics}, change-induced bugs often involve challenging event sequences, diverse test data, and cross-scenario side effects.
To address these challenges, the Generator consists of three subcomponents: \textit{Test Scenario Generation}, \textit{Event Sequence Enhancement}, and \textit{Test Data Enhancement}.
Given a PR, \textit{Test Scenario Generation} performs LLM-based change-impact analysis to identify potentially affected usage scenarios and produces initial test scenarios.
\textit{Event Sequence Enhancement} then retrieves additional application-specific knowledge via targeted queries and uses it to enrich the event sequences of these scenarios.
Finally, \textit{Test Data Enhancement} identifies required test data and constraints, and generates concrete data instances, producing executable test scenarios with realistic test data.

\begin{wrapfigure}{r}{0.6\textwidth}
    \centering
    \vspace{-0.3cm} 
    \setlength{\belowcaptionskip}{-0.2cm}
    \includegraphics[width=0.6\textwidth]{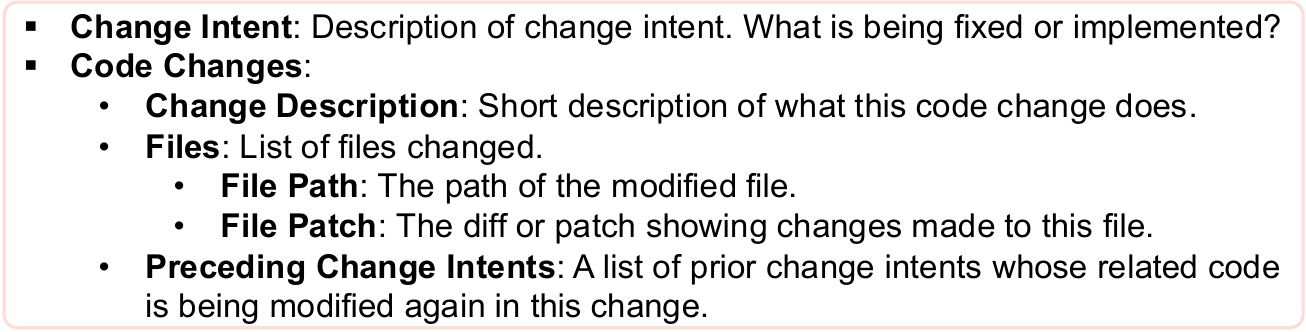}
    \caption{Input Description in Prompt
    }
    \label{fig:input_description}
\end{wrapfigure}
\subsubsection{Input Preparation}

To generate test scenarios that are both aligned with the intended change and effective at revealing unintended bugs, \name{} uses the change intent and code changes of each PR as inputs, as illustrated in Fig~\ref{fig:input_description}.



\textbf{Change Intent.}
For each PR, we collect the PR description as well as any explicitly resolved issue(s), including their summaries and detailed descriptions, if available.

\textbf{Code Changes:}
Capture the concrete implementation details of a PR.

\textit{Change Description and Modified Files:}
Use implementation-level descriptions
(e.g., commit messages) to understand how the change is realized in code.  
Extract fine-grained
change information, including modified files, file paths, and detailed patches.

\textit{Preceding Change Intents.}
As discussed in Section~\ref{subsec:characteristics}, code changes may introduce cross-scenario side effects.
To mitigate this issue, we leverage historical traceability information in software repositories.
For each modified or deleted line in a PR, we use code-commit traceability (e.g., GitHub blame) to identify earlier commits that last modified the same code regions, and retrieve the corresponding PRs and the issues they resolved.
We treat the summaries and descriptions of these historically related PRs and issues as preceding change intents, as the changed code regions were originally introduced or modified to satisfy those intents.
To prioritize relevance, we rank preceding change intents based on their degree of code overlap with the current PR, with higher overlap indicating a higher likelihood of impact.
This design enables \name{} to validate the intended behavior of the current PR while explicitly guarding against regressions affecting previously addressed scenarios.
For example, as illustrated in Section~\ref{subsec:characteristics}, the PR fixing \pinkhref{https://bugzilla.mozilla.org/show_bug.cgi?id=1792881}{Issue~1792881}
 (a scrollbar-related defect) overlaps with code previously modified to support search bar functionality.
As a result, the corresponding search-related issues and PRs are identified as preceding change intents, prompting \name{} to generate test scenarios that verify both the scrollbar fix and the preservation of correct search bar behavior.

\subsubsection{Test Scenario Generation}


Generating test scenarios requires analyzing natural-language change intent together with corresponding code changes, which may span multiple programming languages.
Traditional change impact analysis techniques~\cite{cia_ren2004chianti,cia_ryder2001change} are insufficient in this setting, lacking support for multiple languages.
Accordingly, we use LLMs to perform test scenario generation, leveraging their capabilities in natural language and multi-language code understanding.


\textbf{Initialization.}
We initialize the LLM using a system-level role-play prompt tailored for test scenario generation based on change intent and code changes.
The prompt consists of three components: \textit{Role Assignment}, which defines the LLM as a test scenario generation tool; \textit{Task Description}, which provides step-by-step instructions; and \textit{Output Guidelines}, which specify the required response format, including fields such as ``Change Impact Analysis'' and ``Test Scenarios''.

\textbf{Input Description.}  
As illustrated in Fig~\ref{fig:input_description}, 
the inputs consist of the change intent and its corresponding code changes.  
\textbf{Impact Analysis and Scenario Generation.}
Given the change intent and corresponding code changes, the LLM first analyzes the rationale and desired outcomes of the change and then examines the concrete code modifications that realize this intent.
Based on this understanding, it performs change-impact analysis to identify potentially affected end-user scenarios, including high-risk or sensitive cases.
Finally, it generates test scenarios derived from the impact analysis, covering both new or modified behaviors and critical existing workflows.
All scenarios are written strictly from the end-user perspective, excluding code, internal implementation details, and developer jargon.

\begin{figure*}[t]
    \centering
    \setlength{\abovecaptionskip}{2pt}
    \setlength{\belowcaptionskip}{2pt}

    \begin{minipage}[t]{0.48\textwidth}
        \vspace{0pt} 
        \centering

        \begin{subfigure}[t]{\linewidth}
            \centering
            \setlength{\abovecaptionskip}{1pt}
            \setlength{\belowcaptionskip}{1pt}

            \includegraphics[width=\linewidth]{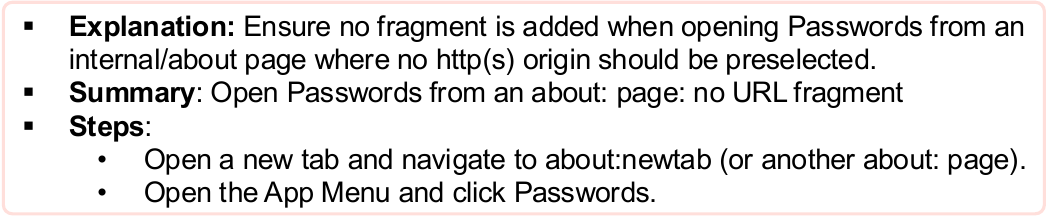}
            \caption{Generated test scenario} 
            \vspace{-2mm}\label{fig:approach_generation_example}
        \end{subfigure}

        \vspace{1.5mm}

        \begin{subfigure}[t]{\linewidth}
            \centering
            \setlength{\abovecaptionskip}{1pt}
            \setlength{\belowcaptionskip}{1pt}
            \includegraphics[width=\linewidth]{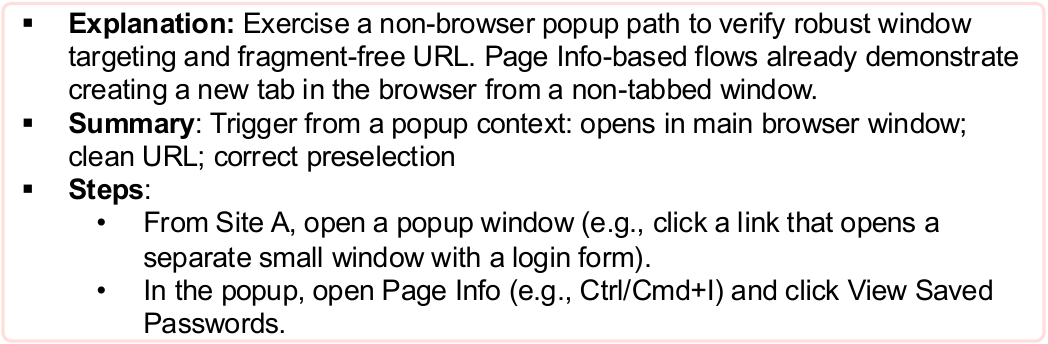}
            \caption{Event-enriched test scenario}\label{fig:approach_path_enhancement_example}
            \vspace{-1mm}
        \end{subfigure}
    \end{minipage}
    \hfill
    \begin{minipage}[t]{0.48\textwidth}
        \vspace{0pt} 
        \centering

        \begin{subfigure}[t]{\linewidth}
            \centering
            \includegraphics[width=\linewidth]{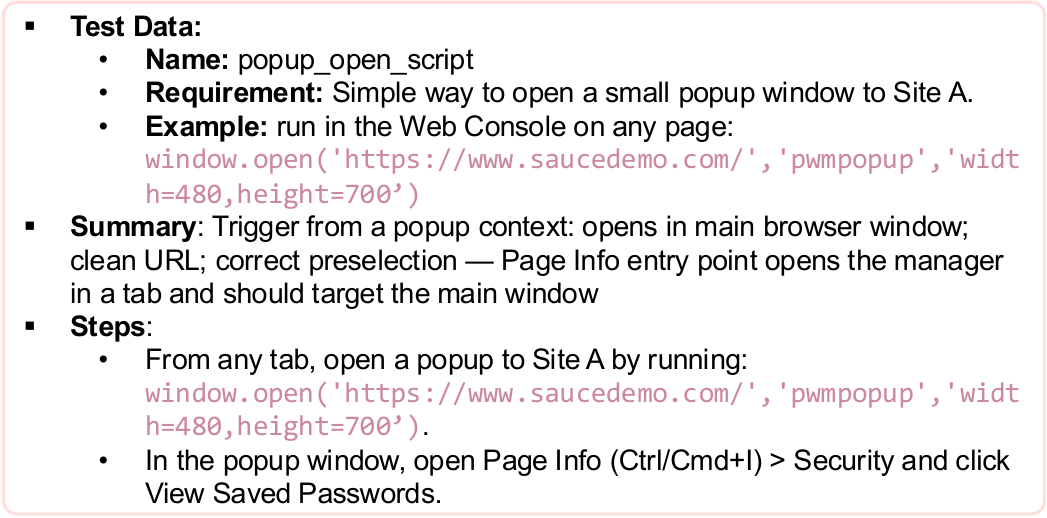}
            \caption{Data-enriched test scenario}\label{fig:approach_data_enhancement_example}
        \end{subfigure}
    \end{minipage}

    \caption{Test Scenario Generation and Enrichment for the PR fixing 
    \pinkhref{https://bugzilla.mozilla.org/show_bug.cgi?id=1858633}{Issue~1858633}.
    }
    \label{fig:approach_scenario_refinement}
        \vspace{-0.6cm}

\end{figure*}

For example, as shown in Fig~\ref{fig:motivation_example}, when testing the PR that fixes
\pinkhref{https://bugzilla.mozilla.org/show_bug.cgi?id=1858633}{Issue~1858633},
\textit{Test Scenario Generation} identifies that the change affects how the Passwords page is opened from the App Menu and other entry points.
Accordingly, \name{} generates test scenarios that exercise this behavior.
One such scenario is shown in Fig.~\ref{fig:approach_generation_example}, verifying that no fragment is appended when opening Passwords page from an internal \texttt{about:} page.


\subsubsection{Event Sequence Enhancement}
After obtaining the initial test scenarios generated by the \textit{Test Scenario Generation}, we further enrich their event sequences to increase event sequence diversity.

\textbf{Scenario Knowledge Base Construction.}
As discussed in Section~\ref{subsec:characteristics}, triggering bugs often requires diverse event sequences.
Past bug reports provide a rich collection of crowdsourced usage scenarios, making them a valuable knowledge base for event sequence enhancement.
For example, 
past bug reports reveal multiple event sequences for opening the Passwords page, such as through the Preferences page, the autocomplete dropdown footer, the Page Info dialog, save/update doorhangers, which are difficult to enumerate manually.
However, not all reports contain meaningful usage scenarios.
To address this, we first apply rule-based heuristics to filter out log-oriented reports (e.g., those with excessive timestamps or keywords such as ``intermittent'').
We then employ LLMs to further identify reports that describe end-user GUI interaction scenarios.
These reports form the \textit{Scenario Knowledge Base} (SKB).
To enable retrieval, we split these reports into overlapping chunks of fixed token size and embed them into a vector space, supporting both keyword-based and semantic search.
During test generation, relevant event sequences are retrieved from the SKB and incorporated via retrieval-augmented generation to enrich event sequence generation.


\textbf{Initialization.}
The LLM is assigned the role of enhancing test scenarios by introducing diverse event sequences based on the change impact analysis.


\textbf{Input Description.}
The change intent explanation, change impact analysis, and test scenarios generated by the \textit{Test Scenario Generation} serve as inputs.
These inputs guide the retrieval of relevant event sequence knowledge, which is used to enrich the event sequences of the test scenarios.



\textbf{Event Sequence Retrieval and Enrichment.}
First, the component identifies opportunities for increasing event sequence diversity by analyzing the change impact analysis and the initial test scenarios.
Second, it generates targeted queries to retrieve alternative event sequences from historical end-user usage and test scenarios in the SKB.
Third, the retrieved event sequences are integrated to enrich the test scenarios, increasing event sequence diversity.
Finally, the enriched scenarios are validated to ensure they remain simple, self-contained, and independent.
If enrichment introduces excessive complexity, new scenarios are generated instead.
All scenarios are written strictly from the end-user perspective, without reference to internal implementation details.

Following the test scenario in Fig.~\ref{fig:approach_generation_example} and the change-impact analysis indicating that the PR affects how the Passwords page is opened from different entry points, \textit{Event Sequence Enhancement} queries the SKB to identify additional access event sequences.
An enriched scenario is shown in Fig.~\ref{fig:approach_path_enhancement_example}, where the Passwords page is opened via a popup window and the Page Info interface instead of the App Menu.
Notably, opening the Passwords page from a popup window is a critical event for triggering the regression reported in
\pinkhref{https://bugzilla.mozilla.org/show_bug.cgi?id=1937085}{Issue~1937085}.
At this stage, the scenario still contains abstract events (e.g., opening a popup window from a website) without concrete test data (e.g., specifying a website that reliably triggers a popup window).

\subsubsection{Test Data Enhancement}

After enriching test scenarios with diverse event sequences, we further identify and instantiate the required test data to make the scenarios executable.

\textbf{Initialization.}
The LLM is assigned the role of identifying and instantiating the test data required by each scenario.

\textbf{Input Description.}
The input consists of test scenarios enriched with diverse event sequences.


\textbf{Test Data Identification and Enrichment.}
For each test scenario, the LLM identifies the required test data by analyzing the scenario and inferring constraints for each data item, and instantiates them into concrete, executable values.
Existing examples, if present, are validated against these constraints and replaced if non-compliant. Otherwise, realistic real-world data are generated.
To increase coverage, representative data variations are introduced without redundancy, including boundary cases and contrasting valid or invalid inputs when meaningful.
Across scenarios, different representative examples are used for the same data type to maximize diversity.
The instantiated data are integrated into the scenarios, which are validated to remain simple, self-contained, and independent.
If enrichment introduces excessive complexity, new scenarios are generated instead.
All scenarios are written strictly from the end-user perspective, without reference to internal implementation details or developer jargon.




Following the scenario in Fig.~\ref{fig:approach_path_enhancement_example}, \textit{Test Data Enhancement} instantiates abstract data requirements into concrete inputs by synthesizing a script that reliably creates a popup context.
As shown in Fig.~\ref{fig:approach_data_enhancement_example}, this concrete test data enables execution of the scenario that triggers the regression in
\pinkhref{https://bugzilla.mozilla.org/show_bug.cgi?id=1937085}{Issue~1937085}, which only manifests when the Passwords page is accessed via a popup window.

\subsection{Test Scenario Executor}\label{sec:executor}

As shown in Fig~\ref{fig:approach_overview}, the Executor executes generated test scenarios on both the pre-change and post-change builds of the SUT.
It consists of two components:
(1) an LLM-based UI Instruction Translation, which translates high-level scenario steps into structured UI instructions based on the current GUI state; and
(2) an algorithmic UI Instruction Execution, which executes these instructions in a containerized environment.
This process is repeated until the test scenario is fully executed or a predefined execution budget is reached.

\subsubsection{UI Instruction Translation}

\textbf{Initialization.}
The \textit{UI Instruction Translation} is initialized with a system-level role-play prompt and maintains a memory session that records previously executed UI instructions.
For each scenario step, it translates the high-level description into structured UI instructions and invokes the \textit{UI Instruction Execution} to execute them in a containerized environment.
We define eleven standard UI actions: click, right-click, long-click, double-click, triple-click, input, scroll, drag, move, keypress, and wait.
Some actions require arguments: click, right-click, long-click, double-click, triple-click, and move require a position; input requires a position and text; scroll requires a direction; and keypress requires key values.

\textbf{Input Description.}
The inputs consist of test scenarios enriched with diverse event sequences and associated test data, together with the current GUI state of the SUT, represented visually (e.g., as a screenshot) to guide UI instruction translation.



\textbf{UI Instruction Translation.}
\textit{UI Instruction Translation} analyzes the test scenario together with the current GUI state (e.g., a screenshot) to determine the next step to execute.
If the step has unmet preconditions, it generates UI instructions to satisfy them before performing the main step, treating precondition actions as regular steps.
It then produces structured UI instructions to carry out the step.
Specifically, UI instructions include the action (e.g., click), the name and position of the target GUI element (coordinates), the input text if the action is an input, the keys if the action is a keypress, and the scroll direction if the action is a scroll.
To improve execution efficiency, when multiple UI instructions can be generated without temporal dependencies, they are batched into a single response.
For example, when saving a login, if the website, username, and password fields as well as the Save button are available, the component generates instructions to fill the fields and click the Save button in one response.

\subsubsection{UI Instruction Execution}

After structured UI instructions are generated, the \textit{UI Instruction Translation} invokes the \textit{UI Instruction Execution} to execute them on the SUT.
The execution component interacts with the GUI using command-line UI utilities (e.g., \texttt{xdotool}) to perform actions such as clicking, inputting, and scrolling.
After execution, it captures a new screenshot to record the updated GUI state, which is used to guide subsequent steps.

All tests are executed within isolated Docker environments to ensure environmental consistency.
For each SUT, we first construct a base Docker image that encapsulates the required build tools, runtime dependencies, and environment configuration.
For each PR under test, we derive a Docker image from this base image, within which we build both the pre-change and post-change versions of the SUT to obtain the corresponding executable files.
Each test scenario is then executed in a fresh container instance for each build, where the container launches the executable of the corresponding SUT version under test.
This setup eliminates interference from residual files, cached state, or configuration artifacts from prior executions.
To ensure fair and consistent comparison, the UI instructions generated for the post-change build are reused verbatim for the pre-change build.

\begin{figure}[!ht]
  \centering
  \begin{subfigure}{0.49\textwidth}
    \centering
    \includegraphics[width=\linewidth]{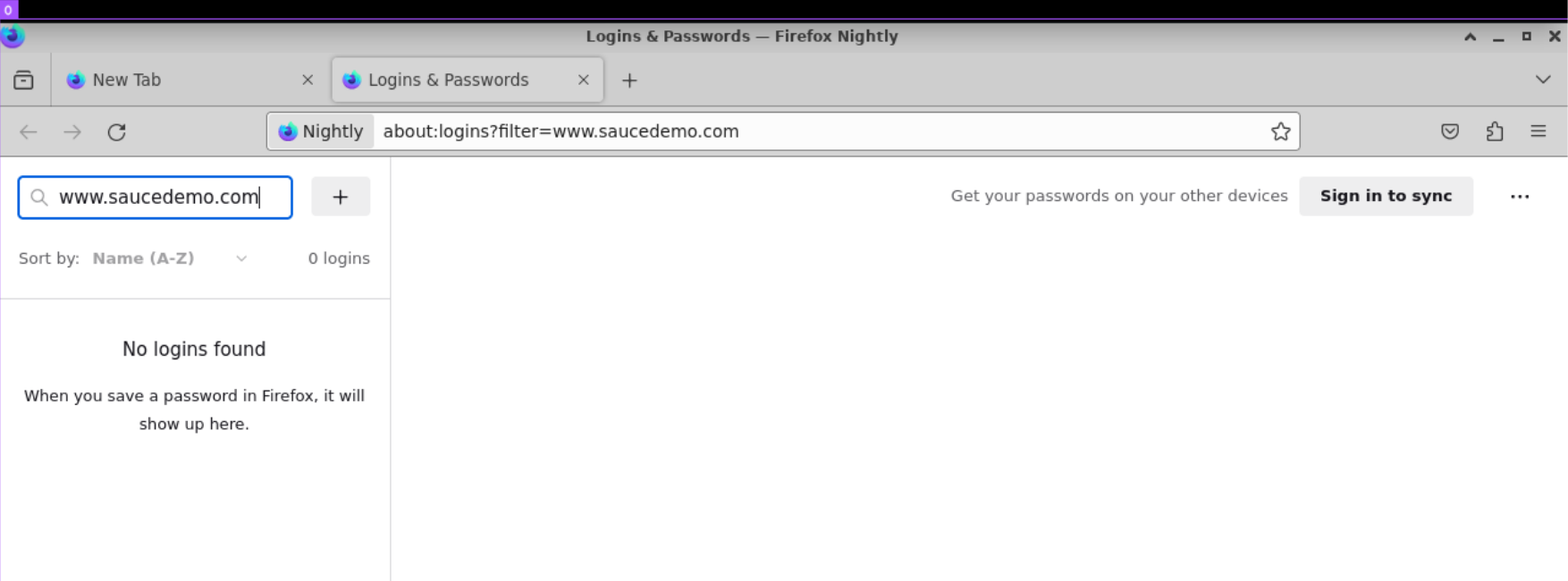}
    \caption{Pre-change Build}
    \label{fig:execution_example1a}
  \end{subfigure}
  \hfill
  \begin{subfigure}{0.49\textwidth}
    \centering
    \includegraphics[width=\linewidth]{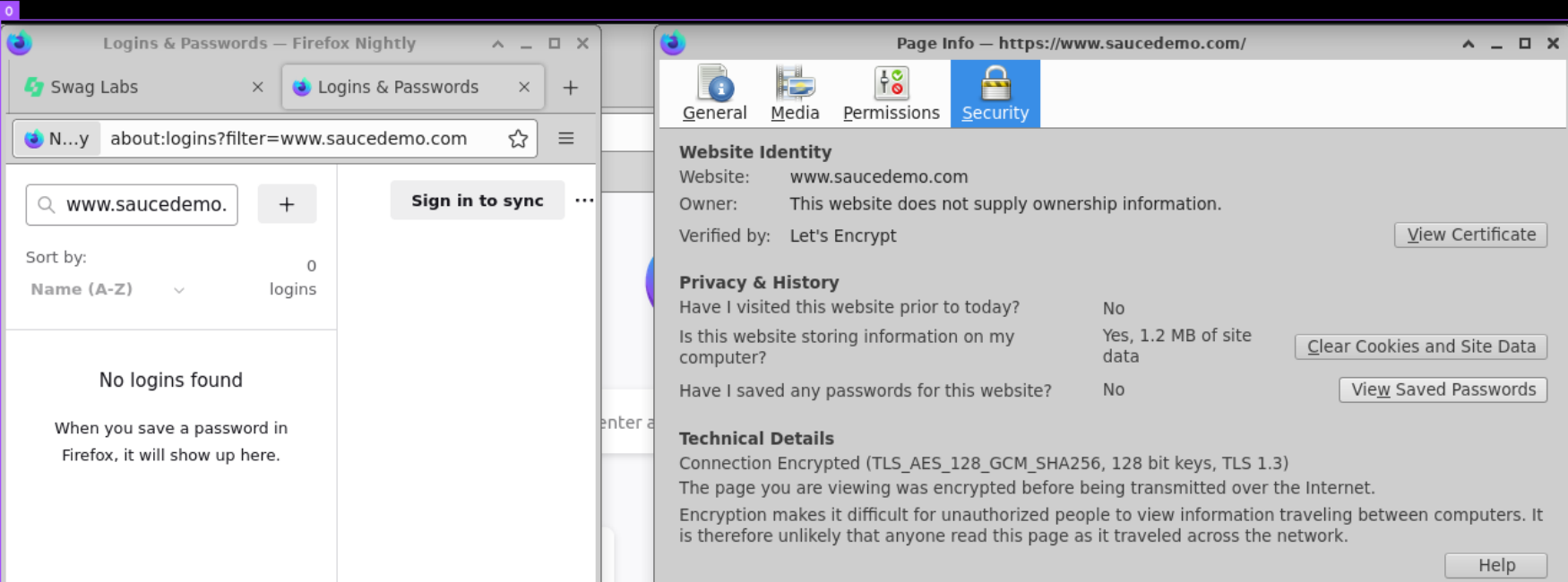}
    \caption{Post-change Build}
    \label{fig:execution_example1b}
  \end{subfigure}
  \caption{Test Scenario Execution for PR Fixing \pinkhref{https://bugzilla.mozilla.org/show_bug.cgi?id=1858633}{Issue1858633} }
  \label{fig:execution_example}
  \vspace{-0.4cm}
\end{figure}

Following the scenario in Fig.~\ref{fig:approach_data_enhancement_example}, the Executor executes the test on both builds.
The Passwords page opens in the main window in the pre-change build (Fig~\ref{fig:execution_example1a}) and in a popup window in the post-change build (Fig~\ref{fig:execution_example1b}), revealing the regression reported in
\pinkhref{https://bugzilla.mozilla.org/show_bug.cgi?id=1937085}{Issue~1937085}.
Although triggered by different event sequences, both executions expose the same underlying defect.

\vspace{-1em}

\subsection{Bug Detector}\label{sec:detector}

After executing test scenarios on both the pre-change and post-change versions of the software, the Detector analyzes the GUI states to identify unintended behavioral changes.
It consists of three components: (1) algorithmic pixel-based GUI difference detection and annotation, (2) LLM-based bug detection, and (3) LLM-based bug filtering.


\subsubsection{Pixel-based Difference Detection and Annotation.}
To automatically identify visual changes between pre-change and post-change GUI states, we first perform pixel-level comparison of the corresponding screenshots to detect differences.
Minor variations are filtered using a threshold (set to 30 in this work), after which morphological dilation is applied to connect nearby difference pixels.
We then extract connected components to identify contiguous regions of change and generate bounding boxes to localize each differing region.
The resulting structured difference information consists of a list of dictionaries, each containing an index and the bounding box coordinates of a detected region, enabling automated identification and localization of visual changes.
Finally, these regions are annotated on the screenshots with labeled bounding boxes, providing a clear visualization of the affected GUI elements.
Both the structured difference information and the annotated GUI screenshots are used as inputs for subsequent bug detection analysis.

\subsubsection{Bug Detection}
\textbf{Initialization.}
\textit{Bug Detection} is initialized with a system-level role-play prompt that assigns the LLM the role of interpreting pre-change and post-change GUI differences to identify unintended issues introduced by a code change.
The prompt treats GUI differences not explicitly described in the change intent as potential bugs, while excluding transient artifacts such as incomplete renderings and out-of-sync GUI states caused by failed or delayed step execution.
A dialogue session is maintained to retain context across detection steps.



\textbf{Input Description.}
The inputs include the change intent and the corresponding change intent explanation, which details the rationale behind the code changes and is generated by \textit{Test Scenario Generation}.
In addition, \textit{Bug Detection} takes the executed test scenario as input.
For each UI instruction executed in a test scenario, the following information is provided: (1) the UI instruction executed, (2) a screenshot of the GUI state before the code change, (3) a screenshot of the GUI state after the code change, and (4) \textit{Parsed Info}, a structured representation of detected GUI differences between the two screenshots.
Each detected difference is assigned an index and localized by a bounding box (\textit{bbox}: $x_1, y_1, x_2, y_2$), and the corresponding numbered bounding boxes are overlaid on the screenshots to visually highlight the detected differences.


\textbf{Bug Detection.}
Given the parsed differences and annotated screenshots for each executed action, \textit{Bug Detection} first performs a holistic GUI analysis to summarize the visible interface state and enumerate present components.
It then conducts a differential analysis by traversing the \textit{Parsed Info} (indexed differences with \textit{bbox}: $x_1,y_1,x_2,y_2$) together with their visual annotations to describe concrete changes, including element additions or removals, position or size shifts, and content or property updates (e.g., labels and values).
Each detected difference is classified as either expected, when it aligns with the change intent and its explanation, or bug, when it constitutes an unintended deviation beyond the specified scope.
Beyond per-difference classification, the \textit{Bug Detection} evaluates the consistency and impact of detected changes on the overall interface, checking for layout misalignment, redundant or missing elements, inconsistent or conflicting states or behaviors, broken or confusing interactions, and potential usability or accessibility degradation.
For each unique unintended issue, the \textit{Bug Detection} produces a concise reasoning trace and files a bug report, avoiding speculation and reporting only clear, observable defects.


\begin{wrapfigure}{r}{0.6\textwidth}
    \centering
    \vspace{-0.4cm}
    \includegraphics[width=0.6\textwidth]{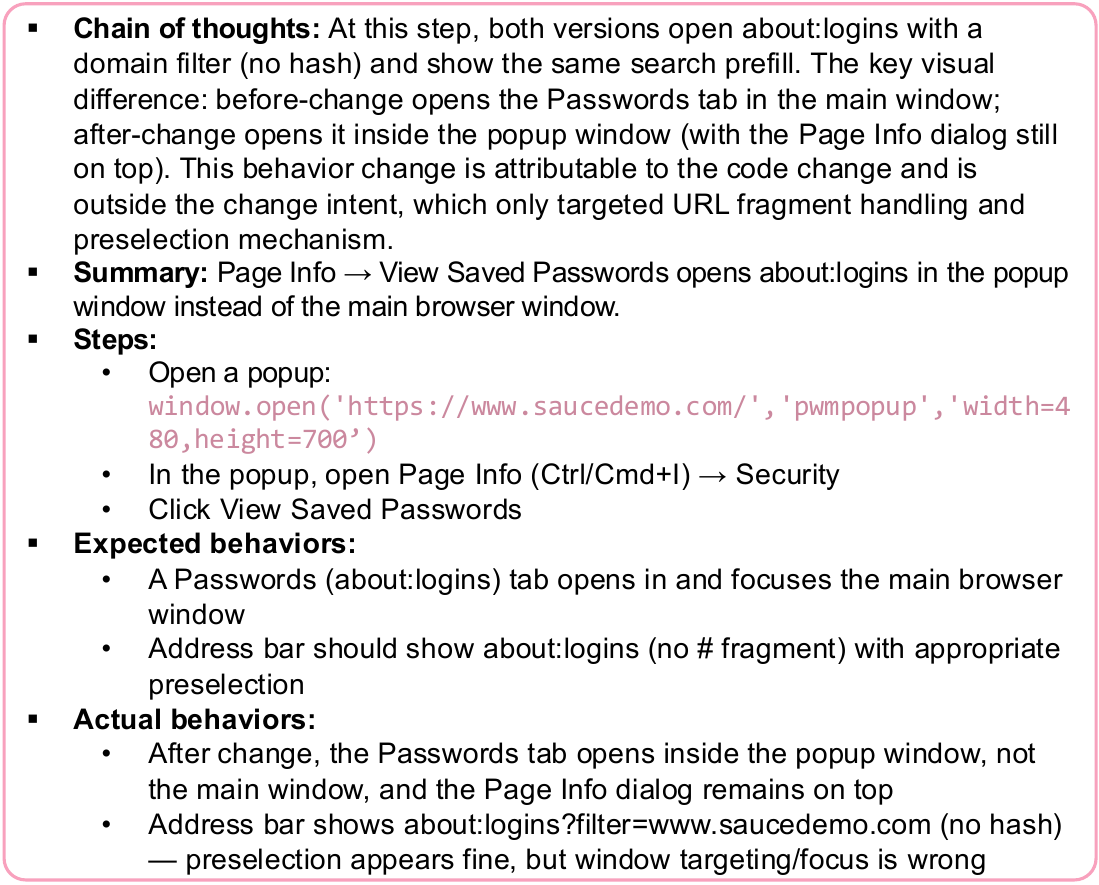}
    \caption{Bug Detection for PR Fixing \pinkhref{https://bugzilla.mozilla.org/show_bug.cgi?id=1858633}{Issue~1858633}}
    \label{fig:approach_detector_example}
        \vspace{-0.4cm}

\end{wrapfigure}
\subsubsection{Bug Filtering}
For each PR under test, \name{} executes multiple test scenarios to maximize coverage.
As a result, the same underlying bug may be triggered repeatedly across different scenarios or multiple times within a single scenario.
In addition, \textit{Bug Detection} may produce spurious reports due to transient rendering artifacts in screenshots.
Consequently, the raw output of \textit{Bug Detection} may contain duplicate or invalid bug reports.
To address this issue, we employ an LLM-based post-processing step to filter bug reports before producing the final results.



\textbf{Duplicate Bug Filtering.}
We identify and remove duplicate bug reports.
Two reports are considered duplicates if they describe the same root cause,
even if they differ in wording.
Among duplicate reports, we retain a single representative instance.
\textbf{Filtering Rendering-Timing Artifacts.}
We filter out false positives caused by screenshot timing or rendering delays.
Because screenshots may be captured before the GUI has fully stabilized, transient visual differences, such as missing caret indicators, partially rendered widgets, or temporarily incorrect layout states, may be mistakenly reported as bugs.
\textbf{Filtering Non-Deterministic GUI Behavior.}
We remove bug reports caused by unstable or non-deterministic GUI behavior.
These include differences arising from layout shifts, random ordering of UI elements, or inconsistent panel selection across executions.
Such behaviors are common in complex GUI applications and do not indicate functional defects.




Following the test scenario execution in Fig.~\ref{fig:execution_example}, the Detector first applies pixel-based difference detection to identify visual discrepancies between the two GUI states.
Because the GUI states differ globally, the entire interface is enclosed within a single purple bounding box (indexed as 0), as shown in Fig.~\ref{fig:execution_example}.
The Detector then analyzes this detected difference in conjunction with the change intent, as illustrated in Fig.~\ref{fig:approach_detector_example}.
Through reasoning over both visual evidence and natural-language intent, it determines that the observed behavioral difference falls outside the intended scope of the change, which targets only URL fragment handling and preselection mechanisms.
Accordingly, the Detector generates a bug report describing this unintended behavior.

\section{Experiment}\label{sec:experiment}



To evaluate the effectiveness and practicality of \name{} in detecting bugs introduced by code changes, 
we formulate the following research questions:

\begin{description}
    \item[RQ1:] How effective is \name{} at detecting previously unknown bugs? 
    \item[RQ2:] What is the precision of the bugs detected by \name{}? 
    \item[RQ3:] What is the recall of previously known regression bugs?
    \item[RQ4:] What are the computational and monetary costs of \name{}?
\end{description}


\subsection{Experimental Setup}



\subsubsection{Implementation}

Our implementation comprises approximately 13.4K lines of Python code and relies on standard Python libraries for UI automation, image processing, and vector-based retrieval, as well as official SDKs for interfacing with LLM services.
To ensure environment consistency and reproducibility, the build of the SUTs and the execution of test scenarios are performed inside \pinkhref{https://www.docker.com/}{Docker} containers.
Our experiments were conducted on three machines with heterogeneous but comparable hardware configurations:
(1) a MacBook Pro equipped with an Apple M4 Pro chip and 24 GB RAM (macOS 26.2),
(2) a MacBook Pro with a 2.0 GHz quad-core Intel Core i5 and 16 GB RAM (macOS 14.1.2), and
(3) a Linux server with an Intel Core i9-14900 CPU (24 cores, 32 threads, up to 5.8 GHz) and 64 GB RAM, running Ubuntu 24.04.1.
We observed consistent results across these machines, suggesting that \name{} is not hardware-dependent.

We employ state-of-the-art LLMs to explore the upper bound of model capability for this task.
Specifically, we use \pinkhref{https://platform.openai.com/docs/models/gpt-5.2}{GPT-5.2} for the Generator and Detector due to its strong reasoning and multimodal capabilities, and \pinkhref{https://platform.claude.com/docs/en/about-claude/models/whats-new-claude-4-5}{Claude Opus 4.5} for the Executor due to its superior performance in grounding and locating GUI elements.
For auxiliary tasks, we use \pinkhref{https://platform.openai.com/docs/models/gpt-5-nano}{GPT-5 Nano} to filter historical bug reports during SKB construction, leveraging its efficiency for lightweight classification, and \pinkhref{https://platform.openai.com/docs/models/text-embedding-3-large}{text-embedding-3-large} to encode the SKB into vector representations for retrieval.



\subsubsection{Software Under Test}

We selected four widely used open-source desktop applications: \pinkhref{https://www.firefox.com/en-US/}{Firefox}, a fast and privacy-focused web browser; 
\pinkhref{https://www.zettlr.com/}{Zettlr}, an all-in-one publication workbench; 
\pinkhref{https://www.jabref.org/}{JabRef}, a graphical Java application for managing BibTeX and BibLaTeX databases; 
and \pinkhref{https://godotengine.org/}{Godot}, a multi-platform 2D and 3D game engine.
These applications were selected based on diversity, popularity and traceability.  
First, the projects span a diverse range of application domains, 
which allows us to assess the generalizability.
Second, all projects are highly popular. 
As of the time of this work, Firefox has over 11k GitHub stars and 6,302 contributors, Zettlr has over 12.4k stars and 151 contributors, JabRef has over 4.2k stars and 874 contributors, and Godot has more than 105k stars with 3,165 contributors. 
Each project has a long development history of at least eight years and demonstrates sustained active developer engagement, ensuring the
representatives of mature and continuously evolving software systems.
Third, all projects maintain public issue tracking systems and version control histories, enabling the traceability of code changes and resolved issues.

\subsubsection{Dataset}
For SKB construction, we crawl 246,665 Firefox bugs, from which 74,625 bugs are retained after filtering.
For Zettlr, we crawl all issues and pull requests with IDs from 1 to 6,035, retaining 2,832 issues and PRs after filtering.
For JabRef, we crawl issues and pull requests with IDs from 1 to 14,554, resulting in 3,256 issues and PRs in the SKB.
For Godot, we crawl issues and pull requests with IDs from 1 to 113,618, of which 35,657 issues and PRs remain after filtering.
To prevent data leakage, when retrieving SKB entries for a given PR under testing, we only consider issues and PRs created prior to the creation time of that PR.

Firefox provides explicit traceability between code changes and the bugs they introduce, making it suitable for evaluating recall on previously known regression bugs.
From Firefox, we randomly select 30 merged PRs, each of which is associated with at least one introduced bug.
For Zettlr, Godot, and JabRef, where such fine-grained traceability is not consistently available, we focus on evaluating the ability of \name{} to detect previously unknown bugs under realistic development settings.
For Zettlr and Godot, we collect 50 recently merged PRs prior to January 2026.
For JabRef, we initially also collected 50 recently merged PRs.
However, after filtering out PRs that are non-functional or unsuitable for testing, the remaining number was insufficient.
Therefore, we extend the collection window and retrieve the 80 recently merged PRs of JabRef prior to January 2026.
After filtering, we retain 38 PRs for Zettlr, 25 PRs for Godot, and 18 PRs for JabRef.

\begin{wraptable}{r}{0.40\textwidth}  
\vspace{-0.3cm}
\caption{Bug Detection}
\setlength{\abovecaptionskip}{0.2pt}  
\setlength{\belowcaptionskip}{-0.6pt}  
\label{tab:bug_finding}
\vspace{-2mm}
\centering
\renewcommand{\arraystretch}{1.1}
\resizebox{0.40\textwidth}{!}{ 
\begin{tabular}{lccccc}
\toprule
\textbf{SUT} & \textbf{PR \#} & \textbf{Bug \#}  & \textbf{TP \#} & \textbf{FP \#} & \textbf{Prec.} \\
\midrule
Firefox  & 30 & 29 & 52             & 41             & 0.559 \\
Zettlr & 38 & 63 & 67 & 95 & 0.414 \\
JabRef & 18 & 17 & 19 & 21 & 0.475 \\
Godot  & 25 & 10 & 10 & 14 & 0.417 \\
\midrule
Total & 111 & 119 & 148 & 171 & 0.464 \\
\bottomrule
\end{tabular}

}
\vspace{-3mm}
\end{wraptable}

\subsection{Effectiveness in Detecting Previously Unknown Bugs (RQ1)}\label{subsec:evaluation_rq1_unknown_bugs}

We manually inspect all bugs reported after testing PRs across all SUTs to determine whether \name{} identifies previously unknown bugs.
As shown in Table~\ref{tab:bug_finding}, 
\name{} detected 119 unique bugs.
Overall, 26 of the detected bugs still exist in the latest project versions.
After reporting these bugs to the corresponding development teams, 16 bugs have already been fixed, two have been confirmed, six remain under discussion, and two were marked as working as intended.
The bug counts in Table~\ref{tab:bug_finding} reflect deduplicated unknown bugs, whereas the true positive (TP) counts include duplicates.
For Firefox, known ground-truth bugs are also excluded for the bug count.

\begin{figure}[!ht]
  \centering
  \begin{subfigure}{0.46\textwidth}
    \centering
    \includegraphics[width=\linewidth]{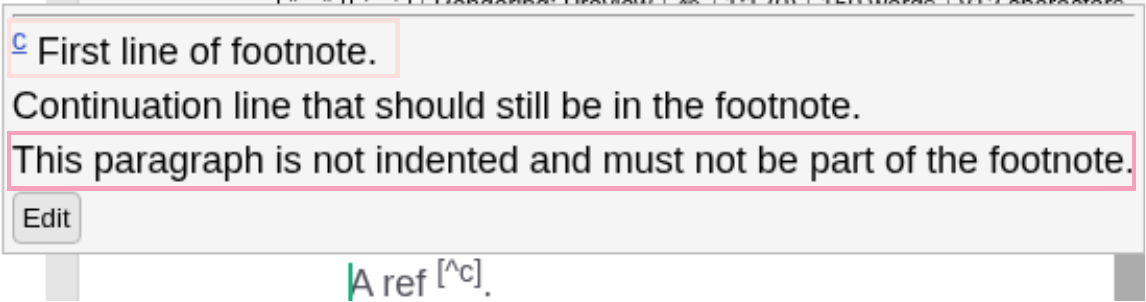}
    \caption{Pre-change Build}
    \label{fig:evaluation_example_a}
  \end{subfigure}
  \hfill
  \begin{subfigure}{0.52\textwidth}
    \centering
    \includegraphics[width=\linewidth]{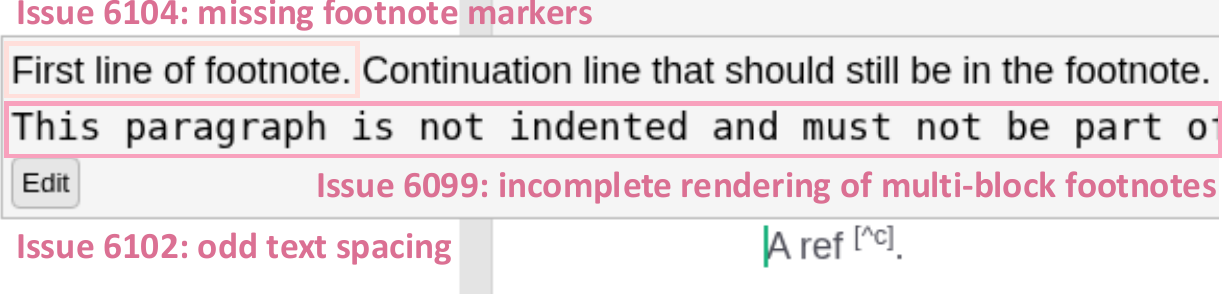}
    \caption{Post-change Build}
    \label{fig:evaluation_example_b}
  \end{subfigure}
  \vspace{-0.25cm} 
\caption{Subset of bugs introduced by \pinkhref{https://github.com/Zettlr/Zettlr/pull/5976}{PR~5976}}
  \label{fig:evaluation_example}
  \vspace{-0.50cm} 
\end{figure}

We further demonstrate the effectiveness through concrete examples.
\pinkhref{https://github.com/JabRef/jabref/pull/14489}{PR~14489}
 in JabRef improves the HTTP server code but introduces a bug: HTTP server endpoint/libraries/demo fails, returning HTTP 500 with HTML and breaking demo library retrieval (\pinkhref{https://github.com/JabRef/jabref/issues/14807}{Issue~14807}). 
After reporting, the bug was promptly fixed.
The detection of this bug illustrates \name{}’s effectiveness in identifying issues in backend server logic and execution.
As another example, \pinkhref{https://github.com/Zettlr/Zettlr/pull/5976}{PR~5976} in Zettlr refactors the parsing and rendering of Markdown footnotes.
\name{}
analyzes the potential impact, and generates test scenarios focusing on high-risk behaviors such as multi-block footnotes.
After execution, \name{} detects multiple bugs,
including incomplete rendering of multi-block footnotes, odd text spacing, incorrect handling of indented continuations, missing footnote markers, and tooltips appearing for invalid footnotes. (\pinkhref{https://github.com/Zettlr/Zettlr/issues/6099}{Issue~6099}, \pinkhref{https://github.com/Zettlr/Zettlr/issues/6102}{Issue~6102}, \pinkhref{https://github.com/Zettlr/Zettlr/issues/6103}{Issue~6103}, \pinkhref{https://github.com/Zettlr/Zettlr/issues/6104}{Issue~6104}, 
\pinkhref{https://github.com/Zettlr/Zettlr/issues/6106}{Issue~6106}).
Fig~\ref{fig:evaluation_example} shows a subset of these bugs.
After reporting,
all issues were fixed, demonstrating the effectiveness in detecting bugs from non-trivial refactoring changes.

We present another example from Godot.
\pinkhref{https://github.com/godotengine/godot/pull/113611}{PR~113611}
fixes missing ``+'' separators in editor tooltips describing keyboard shortcuts, a change that appears purely cosmetic at first glance. 
Through change-impact analysis, \name{} flags potential localization risks, with support from preceding change intents (\pinkhref{https://github.com/godotengine/godot/pull/106943}{PR~106943} and \pinkhref{https://github.com/godotengine/godot/pull/106946}{PR~106946}) that altered overlapping code handling tooltip translation.
Guided by this analysis, \name{} generates and executes test scenarios for tooltip rendering under language-switching conditions. 
It detects a bug (\pinkhref{https://github.com/godotengine/godot/issues/114157}{Issue~114157}
), where switching the editor to French produces mixed-language shortcut tooltips. While the developers traced this to temporarily outdated translations rather than functional code, \name{} successfully detected the user-visible issue and enabled timely translation updates.

In addition, \name{} detects relatively fewer bugs in Godot. 
Closer inspection reveals that this is largely due to Godot’s testing characteristics. Many Godot bugs require constructing a minimal reproduction project before the issue can be triggered. 
Building such projects typically involves non-trivial setup steps, causing test scenarios to frequently exhaust the predefined execution budget before completion.
This limitation could be mitigated by increasing the maximum execution budget. Alternatively, execution efficiency could be improved by reusing existing resources. For example, many Godot PRs and issues already include minimal reproduction projects. Incorporating these projects into the scenario knowledge graph or reusing them through other mechanisms could substantially improve the efficiency and effectiveness of test scenario execution for Godot.



\subsection{Precision of Reported Bugs (RQ2)}\label{subsec:evaluation_precision}

We manually label the issues reported by \name{} for all SUTs as TPs or FPs, and compute precision accordingly.
As shown in Table~\ref{tab:bug_finding}, \name{} detects 111 TP bugs overall, achieving a precision of 46.4\%.
On Firefox, \name{} detects 52 TP bugs from 30 PRs (55.9\%). 
On Zettlr, 67 TP bugs are detected from 38 PRs (41.4\%). 
JabRef yields 19 TP bugs from 18 PRs (47.5\%), while Godot yields 10 TP bugs from 25 PRs (41.7\%).
While \name{} effectively identifies previously unknown bugs across diverse projects, it also incurs a non-negligible false positive rate.


\subsubsection{GUI Rendering and Temporal Instability (78/171: 45.6\%).}
This category covers false positives caused by non-deterministic GUI rendering and variability in screenshot timing due to reliance on screenshots to represent GUI states.

\textbf{Screenshot timing or rendering delay (39 cases).}
The timing of screenshot capture may affect transient visual elements, such as text carets, partially rendered UI components, or delayed dynamic content, leading to benign differences that are mistakenly reported as bugs.

\textbf{Unstable GUI layout or rendering (39 cases).}
These false positives stem from inherent layout variability, where identical actions produce different visual arrangements across executions. 
This is common in Zettlr and JabRef.
For example, 
in Zettlr, opening a file may display the document in different split windows.
In JabRef, the relative ordering of sidebar components, such as the Web Search and Groups panels, may vary across executions.

A potential mitigation is to execute the same UI instructions multiple times and check for consistent GUI states, filtering out transient and non-deterministic visual differences.


\subsubsection{LLM Reasoning and Hallucination Errors (47/171: 27.5\%).}
This category includes false positives caused by limitations in the LLM’s reasoning and grounding.

\textbf{Misinterpreted or hallucinated GUI rendering (38 cases).}
Some false positives arise from how pixel-level differences are annotated in screenshots during detection. Annotated regions may correspond to expected GUI changes, but the LLM can misinterpret them as defects. For example, in one case, a code change intentionally relocates the ``Apply'' button within the ``Clean up entries'' dialog. Although the GUI is correctly updated after the change, the detection annotates the original button location as a differing region, rendering an empty placeholder box in the post-change screenshot. The LLM then interprets this annotated region as a missing or blank button, incorrectly reporting it as a bug.
Other cases involve pure hallucination, where the LLM reports non-existent GUI elements (e.g., an extra edit icon) that do not appear in either screenshot.

\textbf{Assumed execution of unperformed actions (6 cases).}
In these cases, the LLM incorrectly infers GUI states based on actions that were not executed. For example, it reports a missing focus indicator on an input field even though the focus action never occurred.

\textbf{Unnecessary improvement suggestions (3 cases).}
Here, the LLM flags intended behavior as a bug. 
For instance, it reports a disabled ``Rename'' button in Zettlr as an issue, despite this being expected when no renaming action is initiated.

False positives can be mitigated by clarifying the semantics of visual annotations through prompt engineering, preventing the LLM from treating detection markers as GUI elements. For hallucinations caused by incorrect action inference, model fine-tuning with execution-aware data could improve consistency between execution traces and visual reasoning.

\subsubsection{GUI Interaction and Replay Instability (26/171: 15.2\%).}
This category captures false positives caused by non-deterministic GUI interactions and inconsistencies during interaction replay.


\textbf{GUI element localization or interaction failure (15 cases).}
These false positives arise when UI elements are incorrectly located or intended interactions are not executed as planned. 
As a result, the executed interactions deviate from the test scenario, producing GUI states that are mistakenly reported as bugs. 
For example, the LLM reports that pressing Ctrl+A followed by Backspace fails to clear the address field, but inspection shows that the field was never correctly focused. 
In another case, resizing a dialog by dragging fails due to a limitation in our execution implementation, which only specifies the drag start point and relies on a default end point.


\textbf{GUI differences causing replay mismatch (11 cases).}
Here, legitimate GUI changes introduced by the code cause replayed interactions to become misaligned.
UI coordinates that are valid in the post-change build may not correspond to the correct elements in the pre-change build, leading to inconsistent behavior.
For example, 
a code change adjusts the position of the password input field.
During execution on the post-change build, the UI interaction correctly clicks the relocated password field.
However, when the same interaction is replayed on the pre-change build, it fails to target the corresponding field due to the layout difference.
As a result, a popup 
is triggered in the post-change execution, while no such popup appears in the pre-change execution.
False positives can be mitigated by improving the robustness of UI execution (e.g., specifying full drag trajectories), enhancing element localization via execution-aware LLM fine-tuning, and integrating LLMs into the replay phase to detect and adapt to interaction failures.

\subsubsection{Expected GUI Differences due to Code Changes (17/171: 9.9\%).}
The detected visual differences correctly reflect modified behavior or UI states introduced by the PR, but are nonetheless flagged as bugs.
For example, the ``Commit'' button is shown in a disabled (greyed-out) state in the post-change build, whereas it is enabled in the pre-change build.
This difference is an intended result of updated validation logic, which disables the button until specific conditions are met.

False positives can be mitigated through prompt engineering.
The detection prompt can more explicitly emphasize the need to distinguish between unintended bugs and legitimate GUI changes.



\textit{Others (3/171: 1.8\%).}
This small category includes diverse cases: a pre-existing bug present in both pre- and post-change versions, an ambiguous case where domain knowledge is insufficient to classify the bug, and a replay instability issue. 
In the latter, 
the test scenario execution generates a PDF file with randomized file names.
During replay, the execution attempts to reuse the file name generated in the play phase to ensure consistency.
However, the corresponding file does not exist because the replay generates a different file name.
This mismatch leads to file access failures during replay and results in a false positive.

\subsection{Recall on Previously Known Regression Bugs (RQ3)}\label{subsec:evaluation_recall}
\begin{wrapfigure}{r}{0.38\textwidth}
    \centering
    \includegraphics[width=0.38\textwidth]{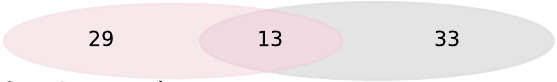}
\caption{Bug Distribution.
}

    \label{fig:venn_firefox_groundtruth}
    \footnotesize
    \raggedright
    \textit{Left:} Unique TP bugs by \name{}. \\
    \textit{Right:} Unique Ground-Truth bugs.
        \vspace{-0.4cm}
\end{wrapfigure}

We focus exclusively on Firefox, leveraging its known ground-truth bugs to assess the extent to which \name{} can discover known bugs.
As shown in Table~\ref{tab:bug_finding}, after testing 30 PRs, \name{} detected 52 TP bugs, among which 10 were duplicates.
After deduplication, 42 unique TP bugs remain.
We further compare the detected bugs with the ground-truth bugs documented in Firefox’s issue tracking system.
Among the 30 PRs, each introduced at least one bug according to the issue tracking records, yielding a total of 54 ground-truth bugs.
Notably, 6 of ground-truth bugs are primarily described through logs or stack traces (e.g., \pinkhref{https://bugzilla.mozilla.org/show_bug.cgi?id=1923374}{Issue~1923374}), making it difficult to precisely infer the corresponding buggy GUI behavior.
As a result, during evaluation, we could not reliably determine whether the bugs detected by \name{} correspond to these log-described ground-truth bugs.
Therefore, we excluded these bugs.
In addition, two bugs marked as duplicates were excluded.
The final ground-truth set used for evaluation contains 46 bugs.
As illustrated in Fig~\ref{fig:venn_firefox_groundtruth}, 13 unique detected bugs by \name{} overlap with the ground-truth bugs, while 33 ground-truth bugs were not detected by \name{}.

\subsubsection{Analysis of Missed Ground-Truth Bugs}
We identified two categories of causes: limitations in test scenario generation coverage and incomplete test scenario execution.

\textbf{Limitations in test scenario generation coverage (20/33: 60.6\%).}
The generated test scenarios lack coverage of the critical trigger steps required to reveal the ground-truth bugs.
These bugs often involve hard-to-predict or strict trigger steps that are difficult to anticipate during test scenario generation.

\textit{Hard-to-predict trigger steps (15/20).}
The PR fixing \pinkhref{https://bugzilla.mozilla.org/show_bug.cgi?id=1745734}{Issue~1745734} (Insecure connection icon is barely visible on HTTP website login forms with the dark theme) introduced \pinkhref{https://bugzilla.mozilla.org/show_bug.cgi?id=1935402}{Issue~1935402}.
Specifically, clicking and dragging the ``This connection is not secure'' panel causes its contents to become invisible.
Such trigger steps involve uncommon interactions (click and drag a panel) and are thus hard for test scenarios to cover.
As a potential improvement, randomization can be introduced into test scenario generation, such as injecting random interaction actions, to increase the unpredictability and diversity of generated test scenarios and thereby improve coverage.

\textit{Strict trigger steps (5/15).}
The PR fixing \pinkhref{https://bugzilla.mozilla.org/show_bug.cgi?id=1592682}{Issue~1592682} introduced \pinkhref{https://bugzilla.mozilla.org/show_bug.cgi?id=1725969}{Issue~1725969}, where the ``Create New Login'' mode is not dismissed after clicking the ``Save'' button when the ``Username'' sort option is active.
The generated test scenarios exercised the sorting functionality and the create-new-login functionality under various preconditions, but did not cover the specific combination where a new login is created while the username sort is active, thereby missing this bug.
In this work, to control the cost of experiments, we impose an upper bound of seven generated test scenarios per PR.
Relaxing this limit and allowing more exhaustive scenario generation could potentially mitigate such missed cases.

\textbf{Incomplete test scenario execution (13/33: 39.4\%).}
Although the test scenarios contain the required trigger steps, the bugs are missed due to incomplete or incorrect execution.

\textit{Reaching the maximum execution budget (4/13).}
The introduced \pinkhref{https://bugzilla.mozilla.org/show_bug.cgi?id=1934682}{Issue~1934682} requires long preconditions to be satisfied.
As a result, the execution of the generated test scenario reaches the predefined maximum execution budget before the trigger steps are exercised, causing the bug to be missed.
Specifically, we impose two execution limits: (1) a maximum LLM interaction budget, which bounds the number of LLM invocations during scenario execution, and (2) a maximum UI instruction budget, which limits the total number of UI interactions executed.
In our experiments, these limits are set to 20 LLM interactions and 35 UI instructions, respectively.
As a potential improvement, increasing the execution budget could help alleviate this issue.

\textit{Incorrect execution (9/13).}
This category consists of five execution-related failure modes that prevent test scenarios from being correctly completed.
(i) \emph{Trigger step execution errors} (4 cases), where interactions required in specific UI components (e.g., password sidebar) are incorrectly executed in other UI contexts (e.g., password page), preventing trigger steps from being properly exercised;
(ii) \emph{execution element location errors} (2 cases), where the execution component interacts with incorrect screen coordinates
(e.g., clicking the ``Create New Login'' button at an incorrect location), causing execution to become stuck;
(iii) \emph{build failures before or after code changes} (1 case), which cause some test scenarios to remain unexecuted;
(iv) \emph{incorrect UI interaction implementation} (1 case), where certain UI interactions (e.g., drag action) require multiple coordinate pairs to specify the drag trajectory, but the current implementation provides only a single coordinate pair,
leading to execution failure; 
and
(v) \emph{Replay failures} (1 case) occur when test scenarios generate randomly named files during execution. 
During replay, the same UI instructions attempt to access these files, which no longer exist due to the random naming. This causes replay failures.

As potential improvements, constructing a more comprehensive scenario knowledge graph that models user tasks with fine-grained interactive steps could better guide test execution.
For other execution-related issues, such as build failures, incorrect UI interaction implementation, and replay failures, more robust engineering solutions could be adopted, e.g., by incorporating LLMs into the replay process to enable adaptive control and error recovery.

\subsection{Computational and Monetary Costs (RQ4)}\label{subsec:evaluation_cost}

\begin{figure}

    \centering
    \includegraphics[width=1\textwidth]{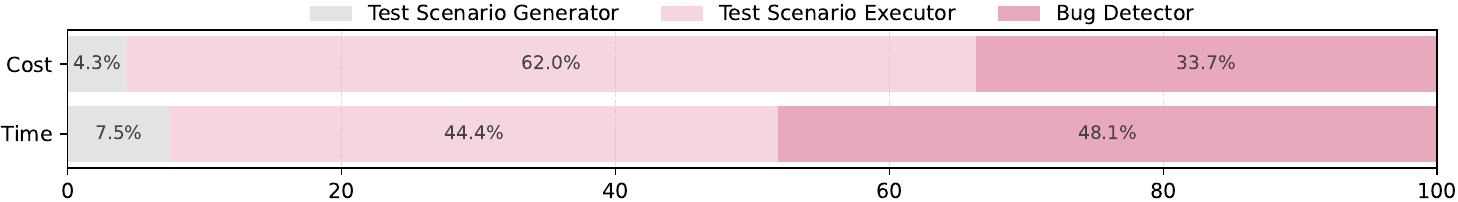}
    \caption{Overhead breakdown: Averages are \$5.99 and 54.8 mins per PR.}
    \label{fig:overhead_breakdown}
        \vspace{-0.4cm}

\end{figure}
SKB construction is a one-off activity.
For Zettlr, it takes 3.22 hours and costs \$0.50.
For JabRef, it takes 5.12 hours and costs \$0.73.
For Godot, it takes 46.21 hours and costs \$5.43.
For Firefox, it takes 50.28 hours and costs \$20.82.
Testing each PR incurs an average time of 54.8 mins and a cost of \$5.99 across all SUTs.
As shown in Fig~\ref{fig:overhead_breakdown}, Generator incurs relatively low time and cost overhead, accounting for 7.5\% of the total execution time and 4.3\% of the total cost, with an average of 4.10 mins and \$0.26 per PR.
Executor constitutes the dominant source of overhead, accounting for 44.4\% of the total time and 62.0\% of the total cost, with an average of 24.3 mins and \$3.72 per PR.
Detector accounts for 48.1\% of the total execution time and 33.7\% of the total cost, with an average of 26.4 mins and \$2.02 per PR.

The relatively high overhead is primarily attributed to Executor and Detector.
Both stages require processing image-based inputs to reason about the current GUI state, generate corresponding UI interaction instructions, and determine whether observed behavioral differences are intended or unintended.
Since our evaluation targets desktop applications, the GUI screenshots are typically of high resolution, resulting in a larger number of image tokens being processed by the model and thus higher computational cost.
In our implementation, we employ state-of-the-art MLLMs (i.e., \pinkhref{https://platform.openai.com/docs/models/gpt-5.2}{GPT-5.2} and \pinkhref{https://platform.claude.com/docs/en/about-claude/models/whats-new-claude-4-5}{Claude Opus~4.5}) to evaluate the approach under strong model capabilities.
This choice allows us to assess the effectiveness of \name{} without confounding factors introduced by weaker models, and provides an upper bound on achievable performance under current LLMs.


Although the time and cost per PR may appear relatively high, it is important to note that \name{} analyzes code changes to identify system-level bugs introduced by the corresponding PRs.
Performing such analysis manually would still require non-trivial developer effort and incur considerable time and labor costs.
Moreover, \name{} complements existing manual testing and code review practices.
Even in mature and rigorously reviewed projects such as Firefox, \name{} is still able to identify bugs that are missed by existing testing and review processes.
By identifying bugs at the PR stage, \name{} can prevent defective code from being merged, thereby improving overall software quality and user satisfaction.
Furthermore, since the detected bugs are directly linked to specific code changes, the cost of fixing them is reduced, as developers can resolve the issues without extensive debugging or localization efforts.

\section{Threats to Validity}\label{sec:threats}
A primary threat is the presence of false positives in the bugs reported by \name{}.
To mitigate this threat, 
we further conducted an in-depth analysis of the main sources of false positives and discussed concrete improvement directions to reduce them in future work.

Our evaluation was conducted on four open-source GUI-based desktop applications, which may not represent all types of software systems.
We mitigate this threat by selecting systems from different domains (web browser, document editor, reference manager, and game engine) with large and actively maintained codebases.
Furthermore, \name{} is not tailored to any specific application or framework, suggesting that the approach can be adapted to other GUI-based systems.
Another threat concerns the choice of LLMs and computational resources.
We used state-of-the-art models to explore the upper bound of achievable performance, and results may differ when using smaller or less capable models.
However, this design choice allows us to evaluate the feasibility and potential of ripple-based GUI testing without conflating methodological limitations with model constraints.
Future work can explore cost–performance trade-offs with alternative models.

\section{Related Work}\label{sec:related}
\textit{Exploratory Testing.}
Exploratory testing has been shown to be effective at the system testing level~\cite{test_computer_software_by_kanerCem,exploratory_nature,exploratory_tester_knowledge,vaaga2002managing,lyndsay2003adventures_session_based_testing,wood2003applying,bach2000session_based_test_management,bach2004exploratory,An_experiment_on_the_effectiveness_efficiency_of_ET}.
A variety of principles and guidelines for conducting and managing exploratory testing have been proposed~\cite{bach2000session_based_test_management,exploratory_book_tips_tricks_tours_techniques_to_guide_test_design,cem2013introduction_to_scenario_testing,soap_opera_testing,lyndsay2003adventures_session_based_testing}.
Several tools and techniques have been developed to support exploratory testing.
Tapir~\cite{tapir} generates navigational test cases to reduce duplication.
Scenario-based exploratory testing, also known as soap opera testing~\cite{soap_opera_testing}, designs complex and realistic usage scenarios to provoke failures that are difficult to anticipate upfront.
Recent work~\cite{syskg,soapoperatg,su2024enhancing} supports soap opera testing by constructing a system knowledge graph that captures user tasks and failures, and by generating test scenarios through combinations of related behaviors.
SoapOperaTG~\cite{soapoperatg} further optimizes the generation and selection of such tests.
Su et al.~\cite{su2025automated} automate the execution of these exploratory test scenarios and associated bug detection with human-like creativity and intelligence.
Despite their strong capabilities in test scenario generation, execution, and bug detection, these approaches are not code-change aware, as they do not leverage information about the specific code changes being introduced.
In contrast, our work is explicitly triggered by a code change and systematically explores potentially impacted scenarios to uncover bugs introduced by that change.

\textit{Change Impact Analysis.}
Change Impact Analysis has been widely studied to reason about the effects of code modifications~\cite{cia_law2003whole,cia_ren2004chianti,cia_ryder2001change,cia_de2008traceability,cia_ying2004predicting,cia_arnold1996software}.
Ryder and Tip~\cite{cia_ryder2001change} characterize change impact in object-oriented programs and highlight how language features such as inheritance and dynamic dispatch lead to non-local effects.
Ren et al. propose Chianti~\cite{cia_ren2004chianti}, a practical CIA tool for Java that identifies affected regression tests by analyzing code-level dependencies between program versions.
These approaches are effective at the program and test level but are largely limited to single programming languages and rely on static or dynamic code analysis.
They do not reason about end-user GUI behavior.
In contrast, our work is change-driven but language-agnostic, and explores the ripple effects of code changes at the level of user-visible GUI interactions.

\textit{Change-aware Regression Testing.}
Our work is also related to regression test selection and prioritization~\cite{regression_elbaum2002test,regression_harrold2002empirical,regression_yoo2012regression}, but fundamentally differs in that we generate new tests tailored to the given code change.
Recent work has proposed techniques that leverage code changes or pull requests to guide regression testing and validation~\cite{chaco_zhou2026change,changeguard2025groninger,testora2026pradel}.
ChaCo~\cite{chaco_zhou2026change} focuses on last-mile regression test augmentation by identifying code changes in a pull request and selecting or augmenting existing tests to improve coverage of the modified code.
The approach operates at the code and test level and aims to improve regression test adequacy for specific changes, but it does not reason about GUI-level effects induced by a change.
ChangeGuard~\cite{changeguard2025groninger} validates code changes via learning-guided execution, using pairwise comparisons between program states before and after a change.
While effective at detecting behavioral differences, the approach primarily targets program-level executions and does not explicitly model end-user interaction scenarios or GUI behaviors.
Testora~\cite{testora2026pradel} leverages natural language information associated with pull requests to classify behavioral differences exposed by generated tests as intended or unintended.
Testora focuses on API-level regression detection in single-language codebases and does not consider GUI-level interactions.
In contrast to these approaches, our work targets GUI-centric software systems and explicitly treats a code change as the epicenter of ripple effects across multiple user-visible scenarios, enabling the discovery of indirect and cross-scenario bugs that are difficult to anticipate at the code or API level.

\textit{Scenario Knowledge Mining.}
Prior work has explored mining execution information from bug reports or issues to reproduce bugs, such as translating steps-to-reproduce into test cases~\cite{automatically_translate_bug_into_test_case,ReCDroid_zhao2019recdroid,sidong2024prompting}, or reconstructing execution traces from bug videos~\cite{bernal2020translating_video_recordings,havranek2021_v2s_translating_video_recordings_tool}.
Some studies extract structured information from bug reports to support exploratory testing~\cite{soap_opera_testing,soapoperatg,syskg,su2025automated}.
BUGINE~\cite{collaborative_bug_finding,li2020bugine_collaborative_bug_finding} recommends related bugs across applications to assist developers during testing.
While these approaches demonstrate the value of crowdsourced artifacts, they do not leverage scenario knowledge to enrich event sequences for testing code changes.

\section{Conclusions}\label{sec:conclusion}
Code changes can introduce bugs despite extensive testing and code review.
Our analysis of real-world software systems shows that a non-trivial fraction of bugs still escape existing tests due to complex event sequences, challenging test data requirements, cross-scenario side effects, and hard-to-predict test oracles.
We present \name{}, a change-driven exploratory GUI testing system that (i) generates and enriches test scenarios targeted at code changes through LLM-based change-impact analysis and retrieval-augmented generation over a Scenario Knowledge Base, (ii) executes these scenarios on both pre-change and post-change builds, and (iii) detects unintended behavioral differences through multi-modal differential analysis.
Our evaluation on real-world software systems demonstrates that \name{} is effective in practice, uncovering 26 previously unknown bugs that persist in the latest versions.
These results highlight the practical value of change-aware exploratory GUI testing in real-world development workflows.
We further analyze observed limitations, including major sources of false positives and missed bugs, and outline concrete mitigation strategies and directions for future improvements.
More broadly, by explicitly linking code changes to observable user behavior, \name{} lays the groundwork for testing interactive systems with increasingly dynamic and evolving user interfaces.

\section*{Data Availability}\label{sec:data}
Replication package with code and dataset: 
\href{https://github.com/RippleTester/RippleGUItesting}
{\textcolor{softpink}{https://github.com/RippleTester/RippleGUItesting}}

\bibliographystyle{ACM-Reference-Format}
\bibliography{acmart}










\end{document}